\documentclass{JHEP}
\usepackage{epsfig}
\preprint{ {\tt hep-th/0007235} }
\newcommand{\be}{\begin{equation}}
\newcommand{\ee}{\end{equation}}
\newcommand{\bea}{\begin{eqnarray}}
\newcommand{\eea}{\end{eqnarray}}
\newcommand{\eq}[1]{(\ref{#1})}
\newcommand{\del}{\partial}

\title{Tachyon condensation in the D0/D4 system}
\author{Justin R. David \\ Department of Physics, University of
California\\Santa Barbara, CA 93106, USA.\\
\email{justin@vulcan.physics.ucsb.edu} }
\abstract{ 
The D0/D4 system with a Neveu-Schwarz B-field in the spatial
directions of the D4-brane has a tachyon in the spectrum of the (0,4)
strings. The tachyon signals the instability of the system to form a
bound state of the D0-brane with the D4-brane. We use the
Wess-Zumino-Witten like open superstring field theory formulated by
Berkovits to study the tachyon potential for this system. The tachyon
potential lies outside the universality class of the D-brane
anti-D-brane system. It is a function of the B-field.
We calculate the tachyon potential at the zeroth
level approximation. The minimum of the tachyon potential in this
case is expected to  reproduce 
the mass defect involved in the formation of the
D0/D4 bound state.
We compare the minimum of the tachyon potential with the mass defect
in three cases. For small values of the B-field we obtain 70\% of the
expected mass defect. 
For large values of the B-field with Pf$(2\pi\alpha'
B) >0$ the potential reduces to that of the D-brane anti-D-brane
reproducing 62\% of the expected mass defect. For large values of the
B-field with Pf$(2\pi\alpha' B) <0$ the minimum of the tachyon
potential gives 25\% of the expected mass defect.
At the tachyon condensate we show that the (0,4) strings decouple from
the low energy dynamics.
}

\begin{document}

\section{Introduction}
The presence of a tachyon in the spectrum of a string theory does not
necessarily imply that the theory is inconsistent. 
In recent times there is
accumulating evidence that the tachyon found in certain open string
theories actually condense to a new stationary point of the potential.
For example,
the tachyon of the bosonic open string theory represents the
instability of the D25-brane to decay to the vacuum. 
The tachyon of a non-BPS brane of Type II string theories also
represents the instability of the non-BPS brane to decay to the vacuum. 

Recent works have shown the existence of a stationary point in the
tachyon potential. 
This potential is computed 
computed in string field theory using level
truncation as an approximation method \cite{kossam}.
The value of the tachyon potential at this stationary point
agrees with the tension of
the unstable branes.
This has been shown to a remarkable degree of
accuracy both for the bosonic open string field theory and for the
superstring field theory on the non-BPS brane in Type II 
string theories \cite{senzwi, wati, moewati, berk2, smerae,
iqbnaq,raszwi}. 
The physics of the solitons in the tachyon potential also seem to be
reproduced in the level truncation approximation scheme \cite{harkra,
kjmt, msz}. Tachyon condensation has recently been studied
in effective field
theories on non-commutative spaces in \cite{gms,kmr,hklm,jmw, gms2}
and its
relationship to string field theory has been discussed in \cite{wit}.

Tachyons are present in other unstable systems.
They can be interpreted as the instability of the system to decay to 
stable systems. 
For all the cases analyzed so far in string field
theory,
the system decays to the vacuum. It would be interesting to find
systems with tachyons, which decay to BPS states.
For example  consider the tachyon
present for the open  strings stretching between
Dp-brane and D(p+2)-branes. 
This tachyonic potential
was studied in first quantized string
theory in \cite{gns}. It represented the instability of the system to
for the system to form a bound state of Dp and D(p+2)-branes.
Another reason to study other systems with tachyon is that one might
be able to find tuning parameter in the tachyon
potential. For the systems studied so far, the tachyon potential is
universal \cite{sen}. 
Thus there is no tuning parameter which can be used to
compare against the level truncation approximation scheme.
In this paper we study  a system with a tachyon
which decays to a BPS state other than the vacuum. There is also a
tuning parameter in the tachyon potential. 

We consider the system of a single D0 and D4-brane with a
Neveu-Schwarz B-field in the spatial directions of the D4-brane. For
generic values of the B-field there is a tachyon in the spectrum of
the strings joining the D0-brane and the D4-brane \cite{seiwit}. 
The tachyon in the spectrum of the $(0,4)$ strings represents the
instability of this system to form a bound state of D0 and D4-brane.
The tachyon is in the
Neveu-Schwarz sector. We construct the tachyon vertex operator and use
the Berkovits formalism of open superstring field theory 
to evaluate the
tachyon potential at the zeroth level.
The tachyon vertex  contains operators
involving matter primaries. Thus the potential falls outside the usual
universality class of unstable branes in superstring theory. 
We show that there is a stationary point in the tachyon potential
which approximates the mass defect in the formation of the D0/D4 bound
state. 
The tachyon is almost on shell for small values of the B-field. Then
the zeroth level approximation for the tachyon potential contribute to
70\% of the mass defect. For large values of the B-field with
with the Paffian, Pf$(2\pi\alpha' B)>0$
we find that the tachyon potential reduces to that of a D0-brane
anti-D0-brane pair. Thus reproducing 62\% of the mass defect. For large
of the B-field with Pf$(2\pi\alpha' B)<0$ 
we find the zeroth level approximation
contributes to 25\% of the mass defect. The reason for this
might be due to the presence of a 
large number of low lying states in the
spectrum of the $(0,4)$ strings for this case \cite{seiwit}.

The unstable D-brane decays to the closed string vacuum. Thus at the
tachyon condensate the open string modes are not dynamical.
We show that the corresponding statement for the D0/D4
bound state is that the $(0,4)$ strings decouple from the dynamics.
To do this we look at the small fluctuations around the tachyon 
condensate. This is facilitated by the description of 
the D0/D4 bound state as an instanton in the
noncommutative $U(1)$ gauge theory of the D4-brane. 
The tachyon instability is the instablity of zero
size instanton in the noncommutative $U(1)$ gauge theory of the
D4-brane. At the tachyon condensate the $(0,4)$ strings become
massive and decouple from the dynamics.
This is seen from the fact that the 
$(0,4)$ strings
correspond to the scale of the instanton via the ADHM construction.
The noncommutative $U(1)$ instanton does not have scale moduli.
Thus at  the level of the moduli space approximation  we can see that 
the $(0,4)$ strings decouple from the 
dynamics at the tachyon condensate.

The organization of this paper is as follows. In Section 2 we review
the D0/D4 system with the B-field in the spatial directions of the 
D4-brane. 
In Section 3 we set up the string field theory of the $(0,4)$ strings 
We write down the vertex operators corresponding to the tachyon and
calculate the tachyon potential. 
In Section 4 we calculate the expected mass defect in the formation of
the D0/D4 bound state and compare it with the value of the stationary
point in the tachyon potential.
In Section 5 we  identify the tachyon condensate as the instanton of
the noncommutative gauge theory on the D4-brane and show the
decoupling of the $(0,4)$ strings. Section 6 contains our conclusions.
Appendix A contains  the correlation function of
twist operators. Appendix B contains details on 
the calculation of the tachyon
potential.

\section{The D0/D4 system}

In this section we review the D0/D4 system with a Neveu Schwarz
B-field along the spatial direction of the D4-brane \cite{seiwit}.
We discuss the mode
expansions of the strings which join the D0-branes and the D4-branes.

Consider a single D0-brane and a single D4-brane of type IIA string
theory in ten dimensions arranged as follows. The D4-brane is extended
along the directions $x^6, x^7, x^8, x^9$. 
The open string spectrum consists of excitations of strings joining
the D0-branes and the D4-branes among themselves and the excitations
of open strings joining the D0-brane and the D4-brane.
We denote 
the open strings joining the D0-brane and the D4-brane as the $(0,4)$
strings. 
Let the string world sheet
co-ordinates of the $(0,4)$ string 
be $X^\mu (\sigma^0, \sigma^1)$. $\mu$ 
runs from $0, \ldots
9$, and $\sigma ^1$ lies between $0$ and $\pi$. We work with Euclidean
world sheet signature. Now  turn on a constant  B-field along
the spatial directions of the D4-brane. We can choose a generic value
of the B-field as given below
\be
\label{bfield}
B_{ij} = \frac{1}{2\pi\alpha'}
\left (
\begin{array}{cccc}
0 &b &0 &0 \\ 
-b &0 &0 &0  \\
0&0&0&b' \\
0&0& -b'& 0 
\end{array}
\right)
\ee
Where $i,j$ runs from $6\ldots 9$. We choose the metric $g_{ij} =
\delta_{ij}$. The boundary conditions of the world sheet co-ordinates
with these moduli turned on is given by
\bea
\left.
\partial_{\sigma^1} X^i + 2\pi i \alpha' B_{ij}\partial_{\sigma^0} 
X^j \right|_{\sigma^1 = \pi} = 0 \\ \nonumber
\partial_{\sigma^0} X^i |_{\sigma^1 = 0} = 0 \\ \nonumber
\partial_{\sigma^0} X^a |_{\sigma^1 = 0 , \sigma^1 = \pi} = 0 
\hbox{ where } a = 1, 2, 3, 4, 5 \\ \nonumber
\partial_{\sigma^1} X^0 |_{\sigma^1 = 0, \sigma^1 = \pi} = 0
\eea
The non-trivial mode expansions arise for the world sheet co-ordinates
$X^i$, It is convenient to define co-ordinates 
\bea
X^+ = X^6+ iX^7 \;\;\;\; X^- = X^6- i X^7 \\ \nonumber
X^{+\prime} = X^6+ iX^7 \;\;\;\; X^{-\prime} = X^6- i X^7 
\eea
The mode expansions of $X^+$ and $X^-$ are given by
\bea
\label{bosemode1}
X^+ &=& i\sqrt{\alpha'} 
\sum_n \left[  \frac{\alpha^+_{n-\nu}}{n-\nu} e^{-(n-\nu)(\sigma^0 +
i\sigma^1)} - \frac{\alpha^+_{n-\nu}}{n-\nu} e^{-(n-\nu) 
(\sigma^0 - i\sigma^1)} \right] \\ \nonumber
X^- &=& i\sqrt{\alpha'} 
\sum_n \left[  \frac{\alpha^-_{n+\nu}}{n+\nu} e^{-(n+\nu)(\sigma^0 +
i\sigma^1)} - \frac{\alpha^-_{n+\nu}}{n+\nu} e^{-(n+\nu)
(\sigma^0 - i\sigma^1)} \right]
\eea
Where 
\be
\label{phase1}
e^{2\pi i \nu} = -\frac{1+ ib}{ 1-ib}, \;\;\; 0\leq\nu<1 
\ee
The only non zero commutation relations are 
\be
[\alpha^{+}_{n-\nu}, \alpha^{-}_{m +\nu} ] = (n-\nu) \delta(n +m)
\ee
Similar mode expansions arise for the world sheet co-ordinates
$X^{+\prime}$ and $X^{-\prime}$. They are given by
\bea
\label{bosemode2}
X^{+\prime} &=& i\sqrt{\alpha'} 
\sum_n \left[  \frac{\alpha^{+\prime}_{n-\nu'}}{n-\nu'} 
e^{-(n-\nu')(\sigma^0 +
i\sigma^1)} - \frac{\alpha^{+\prime}_{n-\nu'}}{n-\nu'} e^{-(n-\nu')
(\sigma^0 - i\sigma^1)} \right] \\ \nonumber
X^{-\prime} &=& i\sqrt{\alpha'} 
\sum_n \left[  \frac{\alpha^{-\prime}_{n+\nu'}  }{n+\nu'} 
e^{-(n+\nu')(\sigma^0 +
i\sigma^1)} - \frac{\alpha^{-\prime}_{n+\nu'}}{n+\nu'} e^{-(n+\nu')
(\sigma^0 - i\sigma^1)} \right]
\eea
Where
\be
\label{phase2}
e^{2\pi i \nu'} = -\frac{1+ib'}{1-ib'} \;\;\; 0\leq\nu'<1
\ee
Similarly the only non zero commutation relations are
\be
[\alpha^{+\prime}_{n-\nu'}, \alpha^{-\prime}_{m +\nu'} ] = 
(n-\nu') \delta(n +m)
\ee

The mode expansions of world sheet superpartners of the bosonic fields
are fixed by supersymmetry. The mode expansions of $\psi^+$ and
$\bar{\psi}^+$ of $X^+$  is given by
\bea
\label{mode1}
\psi^+ &=& -i\sqrt{\alpha'} \sum_n
\psi^+_{n+1/2-\nu}e^{-(n+1/2 -\nu)(\sigma^0
+ i \sigma^1)} \\ \nonumber
\bar{\psi}^+ &=& i\sqrt{\alpha'} \sum_n
\psi^+_{n+1/2-\nu}e^{-(n+1/2 -\nu)(\sigma^0
- i \sigma^1)}
\eea
The mode expansions of the the superpartners of $X^-$ are given by
\bea
\label{mode2}
\psi^- &=& -i\sqrt{\alpha'} \sum_n
\psi^+_{n+1/2+\nu}e^{-(n+1/2 +\nu)(\sigma^0
+ i \sigma^1)} \\ \nonumber
\bar{\psi}^+ &=& i\sqrt{\alpha'} \sum_n
\psi^+_{n+1/2+\nu}e^{-(n+1/2 +\nu)(\sigma^0
- i \sigma^1)}
\eea
We have written the mode expansions in the Neveu-Schwarz sector. The
only non-zero anti-commutation rules are given by
\be
\label{anticom}
\{ \psi^+_{n+1/2 -\nu} , \psi^-_{m+1/2+\nu}  \} = \delta(m +n)
\ee
The mode expansions and the anti-commutation relations
for the superpartners of $X^{\prime-}$ and $X^{\prime +}$ are
given by replacing the variables in \eq{mode1}, \eq{mode2} and
\eq{anticom} by their primed variables.

To show that the $(0, 4)$ strings have a tachyon in their spectrum 
we evaluate the zero point energy. 
The zero point energy for four of the lowest energy states 
in the Neveu-Schwarz sector are given
by
\bea
E_-^+= -\frac{1}{2} + \frac{1}{2} (\nu + \nu')  &\;\;\;&
E_+ ^+=  \frac{1}{2} - \frac{1}{2}( \nu + \nu') 
\\ \nonumber
E_-^- = \frac{\nu -\nu'}{2}  &\;\;\;&
E_+^- = \frac{\nu'-\nu}{2} 
\eea
Out of these four states two of them are projected out by the GSO
projection. For the case of the D0-brane and the D4-brane the states
with zero point energies $E^+_\pm$ are retained \cite{seiwit}.
Thus we see unless
$\nu+\nu'=1$ there is a tachyon in the spectrum of the $(0,
4)$ strings. For $\nu + \nu' < 1$ the tachyon (mass)$^2$ is given by
$E_-^+/(2\alpha')$ 
and for $\nu + \nu' >1$ the tachyon (mass)$^2$ is given by 
$E_+^-/(2\alpha')$.

\section{String field theory of the $(0,4)$ strings}

In this section we will compute the  potential for the tachyon in the
spectrum of the $(0,4)$ strings using string field
theory. We will compute the tachyon potential upto the zeroth level in
the level truncation approximation.
We have seen in section 2 that the tachyon is in the Neveu-Schwarz
sector. Therefore  we can 
use the Wess-Zumino-Witten like open superstring field
theory formulated by Berkovits \cite{berk}
to calculate the tachyon potential.
\subsection{Open Superstring field theory}
We will briefly review open superstring field theory. 
From this
section onwards we work in the units  $\alpha'=1$
The string field theory action is given by
\be
\label{sftaction}
S= \frac{1}{g^2} \langle\langle (e^{-\Phi} Q_{B} e^{\Phi} )
(e^{-\Phi} \eta_{0} e^{\Phi} ) - \int_0^1 dt (e^{-t\Phi} \partial_t
e^{t\Phi} ) \{ (e^{-t\Phi} Q_{B} e^{t\Phi} ), 
(e^{-t\Phi} \eta_{0} e^{t\Phi} ) \} \rangle\rangle, 
\ee
where $\{A, B \} \equiv AB + BA$. 
The open string coupling constant  of the $(0,4)$ strings,
$g$ is related to the closed string coupling constant coupling
constant $g_c$ by $ g^2 = g_c$. The presence of the Neveu-Schwarz
B-field does not change the relationship of the open string coupling
constant to the closed string coupling constant. This is because the
B-field is perpendicular to the common Neumann directions of the
$(0,4)$ strings.
The string field $\hat{\Phi}$ for the $(0,4)$ strings is
given by
\be
\hat{\Phi} = \Phi_+^{(1)}\otimes I\otimes I + \Phi_+^{(2)}\otimes
\sigma_3 \otimes I +
\Phi_-^{(3)} \otimes \sigma_1
\otimes \sigma_1 + \Phi_-^{(4)} \otimes \sigma_2 \otimes \sigma_1
\ee
Where $\Phi_+^{(1)}, \Phi_+^{(2)}$ stand for GSO even open string
vertex operators and $\Phi_-^{(3)}, \Phi_-^{(4)}$  
stand for  GSO  odd open string vertex operators.
These operators have ghost number $0$ and picture number 
$0$ in the combined conformal
field theory of a 
$c=15$ superconformal matter system and the $b, c,
\beta, \gamma$ ghost system with $c=-15$. 
The bosonized ghost fields $\xi, \eta, \phi$ are related to
$\beta, \gamma$ by
\be
\beta= \del\xi e^{-\phi}, \;\;\; \gamma=\eta e^{\phi}
\ee
$\sigma_1, \sigma_2,
\sigma_3$ are $2\times 2$ pauli matrices and I is the $2\times 2$
identity matrix. The first set of matrices are the external Chan-Paton
factors  and 
the second set of matrices are the internal
Chan-Paton factors. $\hat{Q}_{B} = Q_B\otimes I\otimes
\sigma_3$ where $Q_B$ is
given by
\bea
\label{brst}
Q_B &=& \oint dz j_B (z) = \oint dz \{ c(T_m + T_{\xi\eta} + T_\phi +
c\del c b  + \eta e^\phi G_m  - \eta \del\eta e^{2\phi} b \} \\
\nonumber
T_{\xi\eta} &=& \partial \xi \eta \\ \nonumber
T_\phi &=& -\frac{1}{2} \partial \phi \partial \phi - \partial^2 \phi
\eea
$T_m $ is the mater stress tensor and $G_m$ is the matter
superconformal generator.
$\hat{\eta}_0 = \eta_0\otimes I\otimes\sigma_3$ and 
$\eta_0$ is the zeroth mode of the field $\eta$.
The products of operators in $\langle\langle 
\cdots \rangle\rangle$ is defined by
\be
\langle\langle A_1\ldots A_n \rangle\rangle = \langle f_1^{(n)} \circ
A_1(0) \cdots f_n^{ (n)}\circ A_n(0) \rangle
\ee
Here $\langle \cdots \rangle$ denotes the correlation function
evaluated on the disc including the trace over the internal and the
ordinary (external) Chan-Paton factors.
$f\circ A$ denotes the conformal transform of $A$ by $f$. For the case
of the disc the functions $f_k^{(n)}$ are given by
\be
f_k^{(n)} (z) = e^{\frac{2\pi i (k-1)}{n}} \left( \frac{ 1+ iz}{1-iz}
\right) ^{2/n} \hbox{     for    } n\geq 1
\ee

We now expand \eq{sftaction} to compute the tachyon potential to the
zeroth level. For this it is sufficient to retain  only terms upto the
fourth order in the string field
in the expansion of 
\eq{sftaction}. This is given by
\bea
S &=& \frac{1}{2g^2} \langle\langle \frac{1}{2} (\hat{Q}_B \Phi)
(\hat{\eta}_0 \hat{\Phi}) + \frac{1}{6} (\hat{Q}_B \hat{\Phi})
(\hat{\Phi} (\hat{\eta}_0 \Phi) - (\hat{\eta}_0 \hat{\Phi} \hat{\Phi}
) )\\ \nonumber
&+& \frac{1}{24} (\hat{Q}_B \hat{\Phi} ) (\hat{\Phi}^2 (\hat{\eta}_0
\hat{\Phi}) - 2\hat{\Phi} (\hat{\eta}_0\hat{\Phi}) + (\hat{\eta}_0
\hat{\Phi}) \hat{\Phi}^2) \rangle\rangle
\eea

\subsection{The tachyon vertex operator}

We write down the vertex operator corresponding to the tachyon and
the first excited state in the spectrum of the $(0,4)$ strings.
We work in the gauge \cite{bsz}
\be
b_0\hat{\Phi} = 0, \;\;\; \xi_0\hat{\Phi} =0.
\ee
Let us take without loss of generality $\nu + \nu' <1$. Then the
tachyon vertex operator is given by
\be
\label{tachyon}
\hat{T} = \xi c e^{-\phi}(t_+ \Sigma_\nu \Sigma_{\nu '}
 \otimes \sigma_+\otimes
\sigma_1 + t_- 
\Sigma_{-\nu} \Sigma_{-\nu'} 
 \otimes \sigma_- \otimes
\sigma_1 )
\ee
where
\be
\Sigma_\nu \Sigma_{\nu '}
 = \sigma_\nu e^{i\nu H} \sigma_{\nu'} e^{i\nu' H} \;\;\;\;
\Sigma_{-\nu}\Sigma_{-\nu'} = 
\sigma_{-\nu} e^{-i\nu H} \sigma_{-\nu'} e^{-i\nu' H}
\ee
The twist operators $\sigma_{\pm \nu}, \sigma_{\pm \nu'}$ and $H, H'$
are defined in the appendix. 
$\sigma_{+}$ and $\sigma_{-}$ are given by
\be
\sigma_+ =
\left(
\begin{array}{cc}
0 & 1 \\ 0 &0 
\end{array}
\right)
\;\;\;\;\;
\sigma_- =
\left(
\begin{array}{cc}
0 & 0 \\ 1 &0 
\end{array}
\right)
\ee
The first set of matrices in \eq{tachyon} 
stand for the external Chan-Paton indices,
the second set stand for the internal Chan-Paton indices.
The twist operator $\Sigma_\nu \Sigma_{\nu'}$ and the anti-twist
operator $\Sigma_{-\nu} \Sigma_{-\nu'}$ comes with external
Chan-Paton factors $\sigma_+$ and $\sigma_-$ respectively. 
This is because the insertion of $\Sigma_{\nu}\Sigma_{-\nu}$ changes
the boundary conditions to that of the string joining the D0-brane to
the D4-brane. The mode expansion of these strings are given in
\eq{bosemode1}, \eq{bosemode2}, \eq{mode1}, \eq{mode2}. 
While the insertion
of $\Sigma_{-\nu}\Sigma_{-\nu'}$ changes the boundary conditions to
that of the string joining the D4-brane to the D0-brane, that is a
string of opposite orientation.
Note that the conformal 
dimension of the tachyon vertex
agrees with the zero point energy $E_-^+$.

The tachyon vertex operator in \eq{tachyon} reduces to
the tachyon vertex of a D-brane anti-D-brane pair when $\nu=0$ and
$\nu'=0$. 
When $\nu=0$ and $\nu'=0$ we see from the mode expansions in
\eq{bosemode1}, \eq{bosemode2}, \eq{mode1}, \eq{mode2} the boundary
condition is Dirichlet at both ends of the $(0,4)$ strings. From
\eq{phase1} and \eq{phase2} we see that
$\nu=0$ and $\nu'=0$ can occur only if $b\rightarrow\infty$ and
$b'\rightarrow\infty$ or $b\rightarrow\-\infty$ and
$b'\rightarrow\-\infty$. The D0-brane charge induced on the D4-brane
is proportional to $- \hbox{Pf}(2\pi B) = -bb'$. The
D4-brane effectively becomes an anti-D0-brane. Thus the tachyon vertex
in \eq{tachyon} reduces to the tachyon vertex of a D-brane
anti-D-brane pair. 

For completeness we include the vertex operator of the first excited
state. The first excited state becomes the tachyon and the tachyon in
\eq{tachyon} becomes the first excited state when $\nu+\nu'>1$.
The vertex operator for the first excited state is given by
\be
\hat{E}  
 =  \xi c e^{-\phi} ( e_+ \Sigma_{1-\nu}\Sigma_{ 1-\nu'}
\otimes \sigma_+\otimes
\sigma_1 + e_- 
\Sigma_{-(1-\nu)}\Sigma_{-(1- \nu')}
 \otimes \sigma_- \otimes
\sigma_1)
\ee
where
\bea
\Sigma_{ 1-\nu} \Sigma_{1-\nu'} &=& 
\sigma_{1-\nu} e^{i(1-\nu) H} \sigma_{1-\nu'}
e^{i(1-\nu')H'} \\ \nonumber
\Sigma_{ -(1-\nu)} \Sigma_{-(1-\nu')}
&=& \sigma_{-(1-\nu)} e^{-i(1-\nu) H} \sigma_{-(1-\nu')}
e^{-i(1-\nu')H'} 
\eea
To compute the tachyon potential to the zeroth level we take the
string field  to be
\be
\hat{\Phi} = \hat{T} + \hat{E}
\ee

\subsection{The tachyon potential}

In the case of the unstable brane or the D-brane anti-D-brane system
the tachyon potential is universal \cite{sen}. The tachyon potential 
in this case is independent of the background conformal field theory
except for an overall multiplicative factor. 
\footnote{The overall factor is the open string
coupling constant. The relationship of the open string coupling
constant to the closed string coupling constant can depend on the
background conformal field theory.} 
The computation of the tachyon potential for the 
case involves
correlations functions involving the ghosts fields and the matter
energy momentum tensor with central charge $15$.
It is clear from the vertex operator for the tachyon in \eq{tachyon}
that it involves matter primaries. Thus the tachyon potential 
for the $(0,4)$ strings does not
belong to the same universality class as that of the unstable D-branes.
In fact it depends on the background B-field. 

The details on the computation of the tachyon potential is given in
appendix B. We state the result.
The tachyon potential of the $(0,4)$ strings at
the zeroth level is given by
\bea
V &=& -S = \frac{1}{g^2} \left( 
-\frac{1}{2} (1- (\nu + \nu')t_- t_+  + 
+\frac{1}{2} (1- (\nu + \nu')e_- e_+   \right. \\ \nonumber
&+& \frac{1}{F(\nu, 1-\nu, 1; 1/2)F(\nu', 1-\nu', 1; 1/2)}
t_- ^2 t_+^2   \\ \nonumber
&+& \left. 
\frac{e_-^2 e_+^2}{F(1-\nu, \nu, 1; 1/2) F(1-\nu' , \nu', 1;1/2)} 
+  \frac{2t_+ t_- e_+e_-}{F(\nu, 1-\nu, 1; -1) F(\nu', 1-\nu', 1; -1)}
\right)
\eea
where the hypergeometric function $F(\nu, 1-\nu, 1; x)$ is defined
in \eq{hyper}.  We have the relation \cite{abst}
\be
F(a, 1-a, 1; 1/2) = \sqrt{\pi} \frac{1}{\Gamma (\frac{1}{2} + 
\frac{a}{2} ) \Gamma (1 -\frac{a}{2})} 
\ee
Using this, the tachyon potential becomes
\bea
\label{tpot}
V&=& \frac{1}{g^2} \left( 
-\frac{1}{2} (1-(\nu + \nu')) t_- t_+ 
+\frac{1}{2} (1-(\nu + \nu')) e_- e_+ 
\right. \\
\nonumber
&+&  \frac{1}{\pi}  
\Gamma (\frac{1}{2} + \frac{\nu}{2} ) \Gamma (1-\frac{\nu}{2})
\Gamma (\frac{1}{2} + \frac{\nu'}{2} ) \Gamma (1-\frac{\nu'}{2})
( t_- ^2 t_+^2  + e_-^2 e_+^2 )  \\ \nonumber
&+& \left. 
 \frac{2t_+ t_- e_+e_-}{F(\nu, 1-\nu, 1; -1) F(\nu', 1-\nu', 1; -1)}
\right)
\eea

\section{Analysis of the tachyon potential}

The mass defect for  D-brane anti-D-brane annihilation is given by
$-2 \times$(mass of the D-brane). 
There is a minimum of the tachyon potential
computed by the level truncation approximation scheme which
approximates the mass defect. 
For the D0/D4 system with the Neveu-Schwarz B-field in the spatial
direction we expect the tachyon condensate to be the bound state of
the D0-brane within the D4-brane. 
Thus we expect the minimum of the tachyon potential of the $(0,4)$
strings to approximate the mass defect of the formation of the D0/D4
bound state from the D0-brane and the D4-brane.
In this section we will compare this mass defect 
with the value of the minimum of the tachyon potential.

\subsection{The calculation of the mass defect}

The mass of the D0-brane in the conventions of \cite{sen, bsz} 
is given by
\be
M_{D0} = \frac{1}{2\pi^2g^2}
\ee
This formula is true for any Dp-brane.
In calculating this mass the
longitudinal directions of the Dp-brane is assumed to be compact.
The mass of the D4-brane with the B-field in the spatial directions is
given by
\be
\label{D4}
M_{D4} = \frac{1}{2\pi^2g^2} \sqrt{(1+ b^2)(1+b^{\prime 2})}
\ee
We can understand this expression from the Dirac-Born-Infeld action.
\be
S= \frac{1}{2\pi^2g^2} \sqrt{\hbox{Det}(G +2\pi B)}
\ee
where $G$ is the induced metric. Substituting the value of the B-field
in \eq{bfield} and the metric $g_{ij} =\delta_{ij}$ 
and expanding the Dirac-Born Infeld action 
in the static gauge we obtain the mass for
the D4-brane with the B-field as given in \eq{D4}.
It is instructive to  understand the formula in \eq{D4} by another 
method. 
The BPS formula for the D0/D4 system with moduli is given by
\cite{obpi}
\be
M^2 = (Q_0 + (1- \hbox{Pf}(2\pi B))Q_4 )^2 + (2\pi^2
B_{ik}g^{ij} g^{jl}B_{kl} + 2 \hbox{Pf}(2\pi B)) Q_4^2
\ee
Here $Q_0 = N_0/(2\pi^2g^2)$ and $Q_4 = N_4/(2\pi^2g^2)$. $N_0$ and
$N_4$ are the number of D0-branes and D4-branes respectively.
Substituting the values of the B-field from \eq{bfield} we get
\be
\label{mass}
M^2 = (Q_0 + (1-bb')Q_4)^2 + (b+b')^2 Q_4^2
\ee
Substituting $Q_0=0$ and $Q_4 = 1/(2\pi^2g^2)$ 
in \eq{mass} we get \eq{D4}. 

Now it is easy to write down the mass formula for the D0/D4 bound
state. Substitute $Q_0= 1/(2\pi^2g^2)$ and $Q_4 = 1/(2\pi^2g^2)$
in \eq{mass} we get the following mass formula for the D0/D4 bound
state.
\bea
M_{D0/D4} &=& \frac{1}{2\pi^2g^2}\sqrt{(1+(1-bb'))^2 + (b+b')^2}
\\ \nonumber
&=& \frac{1}{2\pi^2g^2} \sqrt{4 + b^2 + b^{\prime 2} - 2 bb' +
b^2b^{\prime 2} }
\eea
Thus the mass defect is given by
\be
\Delta M = M_{D0/D4} - (M_{D0} + M_{D4})
\ee
We analyze the mass defect $\Delta M$ in three limiting cases.

\noindent
{\bf Case 1.} $b\ll 1, b'\ll1$ \\
The leading order terms in the mass defect is given by
\be
\label{case1}
\Delta M_1 = -\frac{1}{2\pi^2g^2} (\frac{b+b'}{2})^2
\ee

\noindent
{\bf Case 2.}
$b\rightarrow \infty, b'\rightarrow\infty$ or $b\rightarrow
-\infty, b'\rightarrow -\infty$; Pf$(2\pi B)>0$\\
The leading order terms in the mass defect is given by
\be
\label{case2}
\Delta M_2 = -2 \frac{1}{2\pi^2g^2}
\ee
Notice that in this case the mass defect reduces to that of the
D-brane anti-D-brane system. This is expected from the discussion in
Section 3.2

\noindent
{\bf Case 3.} 
\label{case3}
$b\rightarrow\infty, b'\rightarrow -\infty$ or $b\rightarrow
-\infty, b'\rightarrow \infty$; Pf$(2\pi B)<0$ \\
For this case the leading order terms in the mass defect is given by
\be
\Delta M_3 = -\frac{1}{4\pi^2g^2} (\frac{1}{b} + \frac{1}{b'})^2
\ee

\subsection{The minimum of the tachyon potential and the mass defect}

In this section we compute the minimum of the tachyon potential and
compare it with the mass defect obtained in section 4.1.
The tachyon potential in \eq{tpot} has an extrema at
\bea
t_-t_+ &=& \frac{\pi}{4}\frac{1-(\nu+\nu')}{\Gamma(1/2
+\nu/2)\Gamma(1-\nu/2)\Gamma(1/2 +\nu'/2)\Gamma(1-\nu'/2)} \\
\nonumber
e_-=0 &\;& e_+=0
\eea
The minimum of the tachyon potential at this extrema is  given by
\be
\label{tacmin}
V_{{\rm min}} = -\frac{\pi}{16g^2}
\frac{(1-(\nu+\nu'))^2}{\Gamma(1/2+\nu/2) \Gamma(1-\nu/2)
\Gamma(1/2+\nu'/2)\Gamma(1-\nu'/2)}
\ee
We now compare this to the mass defect obtained in the three cases in
section 4.1

\noindent
{\bf Case 1.} $b\ll 1, b'\ll1$ \\
To the leading order the value of $\nu$ and $\nu'$ are given by
\be
\nu = \frac{1}{2} - \frac{b}{\pi}\;\;\;\; \nu' = \frac{1}{2}
-\frac{b'}{\pi}
\ee
We substitute  this values of $\nu$ and $\nu'$ in \eq{tacmin}.
We find the 
minimum of the tachyon to the leading order is given by
\be
V_{{\rm min}} = -\frac{1}{2g^2\pi^2} \left(\frac{b+b'}{2}\right)^2
\frac{\pi}{2(\Gamma(3/4))^4}
\ee
On comparison with the mass defect for this case from \eq{case1} we
obtain $70\%$ of the expected result.
\be
\frac{V_{{\rm min}}}{\Delta M_1} = \frac{\pi}{2 (\Gamma(3/4)^4} = .70
\ee

\noindent
{\bf Case 2.}
$b\rightarrow \infty, b'\rightarrow\infty$ or $b\rightarrow
-\infty, b'\rightarrow -\infty$; Pf$(2\pi B)>0$ \\
As we have seen in section 3.2 this  
case is expected to reduce that of the D0-brane anti-D0-brane
system. Without loss of generality we take $b\rightarrow -\infty,
b'\rightarrow -\infty$. The value of $\nu$ and $\nu'$ to the leading
order are given by
\be
\nu = -\frac{1}{\pi b} \;\;\;\; \nu' = -\frac{1}{\pi b'}
\ee
It is easy to see the tachyon potential in \eq{tpot} reduces to that
of the D0-brane anti-D0-brane system. The minimum of the tachyon
potential is given by
\be
V_{{\rm min}} = -\frac{1}{16g^2} 
\ee
This is the value of the minimum of the tachyon potential computed in
\cite{berk2, bsz} for the D-brane anti-D-brane system. Comparison with
the mass defect in \eq{case2} gives 62\% of the expected result.
\be
\frac{V_{{\rm min} } }{\Delta M_2} = \frac{\pi^2}{16} = .62
\ee

\noindent
{\bf Case 3.} 
$b\rightarrow\infty, b'\rightarrow -\infty$ or $b\rightarrow
-\infty, b'\rightarrow \infty$; Pf$(2\pi B)<0$ \\
Without loss of generality we can consider the case
$b\rightarrow\infty, b'\rightarrow -\infty$. For this case the values
of $\nu$ and $\nu'$ to the leading order are given by
\be
\nu = 1-\frac{1}{\pi b} \;\;\;\; \nu' = -\frac{1}{\pi b'}
\ee
Substituting this in \eq{tacmin} we obtain
\be
V_{{\rm min}} = -\frac{1}{4\pi^2 g^2} \left( \frac{1}{b} + \frac{1}{b'}
\right)^2 \frac{1}{4}
\ee
Comparing the value of the minimum of the tachyon potential 
to the
mass defect in \eq{case3} yields only 25\% of the expected result.
\be
\frac{V_{{\rm min}} }{\Delta M_2} = \frac{1}{4} = .25
\ee
Perhaps the low contribution from the zeroth level approximation to
the tachyon potential might be due to the large number of low 
lying states found for this case \cite{seiwit}.

It is nice to see that we obtain the same dependence of the
mass defect on the B-field in Case 1 and Case 2 
from two different functions.
Graphically it can be easily seen that $V_{\rm{min}} \leq
\Delta{M}$ for all values of $\nu$ and $\nu'$ between $0$ and $1$.
Equality holds only when $\nu+\nu'=1$, that is when the tachyon is
massless and there is no mass defect. The best approximation to the
mass defect over all the allowed values of $\nu$ and $\nu'$ is 70\%.

\section{Excitations around the tachyon condensate}

In this section we examine the excitations of the $(0,4)$ strings
around the D0/D4 bound state.
This is facilitated by the description of the D0/D4 bound state from the
gauge theory of the D4-brane. The existence of the bound state was
shown from the gauge theory in \cite{dmwy}.
The bound state of the D0-brane within the D4-brane 
can be identified with a single  instanton of the D4-brane gauge
theory. We will briefly review the arguments here.

The gauge theory of the D4-brane with the B-field  is a noncommutative
$U(1)$ gauge theory with $N=4$ supersymmetry. This theory admits
non-singular instanton solutions \cite{shnek}. This noncommutative
$U(1)$ gauge theory with a single instanton has the right properties 
to be identified as the bound state the D0-brane within the D4-brane.
The arguments for this is the same as the case with no B-field.
The instanton in the D4-brane gauge theory  carries the
Ramond-Ramond charge of the D0-brane. It preserves the same
supercharges as the D0-brane \cite{seiwit}. The instanton of the
D4-brane gauge theory has the right action as the tension of the
D0-brane. The ADHM construction of the instanton of noncommutative 
D4-brane gauge theory reproduces the moduli space of the D0-brane gauge
theory \cite{shnek}.

\subsection{The decoupling of the $(0,4)$ strings}

In this section we show that at the tachyon
condensate the $(0,4)$ strings become massive and decouple from the
dynamics. 

We examine the small fluctuations around the tachyon condensate. 
The tachyon condensate is identified
with the instanton of the non-commutative gauge theory. 
The small fluctuations around the instanton can be determined by
examining the moduli space of the  instanton of the noncommutative 
$U(1)$ gauge theory.
This is given by the  ADHM construction of the instanton. 
The moduli space is determined by solving the following equations
\bea
\label{moduli}
AB &=& 0 \\ \nonumber
|A|^2 - |B|^2 &=& r
\eea
These are the D-terms of a  $(4,4)$ supersymmetric $U(1)$ gauge theory
with one hypermultiplet, that is the gauge theory of the D0-brane. 
$r$ is the Fayet-Illiopoulous term. It has been shown that a bound
state exists for $r>0$ \cite{dmwy}. 
For Case 1 and Case 2 $r$ has been identified
to be proportional to the square of the self dual part of the B-field
\cite{seiwit}.
$A$  and $B$ are the bosonic components of the hypermultiplet. $A$ is
is a complex field which carries a charge of  $+1$  under the $U(1)$
gauge symmetry and $B$ carries a charge of $-1$ under the $U(1)$
gauge symmetry. To determine the  moduli space we must solve the
equations in \eq{moduli} and identify gauge equivalent configurations.

The solution of \eq{moduli} is given by $B=0$ with $|A| = \sqrt{r}$. 
Using the $U(1)$ gauge degree of freedom one can choose the solution 
$B=0$ and $A=\sqrt{r}$. Thus we see there is no modulus for the
non-commutative instanton other than its position.

Small fluctuations of the $(0,4)$ strings can be identified with the
small fluctuations of the fields $A$ and $B$. 
This can be seen from the fact that the fields corresponding to the
$(0,4)$
strings are hypermultiplets in the gauge theory of the D4-brane and the
D0-brane. 
Therefore at the tachyon condensate, the $(0,4)$ 
strings are massive and
decouple from the low energy dynamics. 
The decoupling of the
$(0,4)$ strings is the analog of the decoupling of the strings joining
the D-brane with the anti-D-brane. It is the analog of the decoupling
of the relative $U(1)$ for the D-brane, anti-D-brane case.

\section{Conclusions}

We have studied tachyon condensation in the D0/D4 system with B-field
in the spatial directions of the D4-brane using string field theory.
The tachyon in the spectrum of the $(0,4)$ strings signals the
instability of the D0/D4 system to form the D0/D4 bound state. Here
after tachyon condensation we are left with a BPS state.
We computed the tachyon potential in the zeroth level approximation.
This tachyon potential is outside the universality class analyzed for
the D-brane anti-D-brane systems. It is a function of the background
B-field.
We compared the minimum of the tachyon potential to the mass defect 
for the formation of the D0/D4 bound state in three cases.
When the tachyon is almost on shell for small values
of the B-field the zeroth level approximation for the tachyon
potential contributes to 70\% of the expected mass defect. For large
values of the B-field with Pf$(2\pi B)$ the tachyon potential reduces
to that of the D-brane anti-D-brane pair. For large values of the
B-field with Pf$(2\pi B)$ the zeroth level approximation contributes
only to 25\% of the mass defect.  Note that in
this last case the level truncation method is not very successful at
the zeroth level. It would be interesting to know whether the 
inclusion of
higher levels for this case converges to the expected answer.

We see that for the case of large values of the B-field with Pf$(2\pi B)
>0$ the D0/D4
system reduces to that of the D-brane anti-D-brane system. 
One is tempted to think that
the decoupling of the open string modes 
in tachyon condensation of the D-brane anti-D-brane system
can be understood from the description of the D0/D4 bound state as an
instanton in the noncommutative gauge theory of the D4-brane for large
values of the B-field.

\acknowledgments
The author thanks Shinji Hirano, Nissan Itzhaki and 
especially Shiraz Minwalla for discussions. He is grateful for
encouragement from Joe Polchinski and Ashoke Sen. The work of the
author is supported by NSF grant PHY97-2202.

\appendix
\section{The twist operator and its correlation function}

This appendix discusses the definition of the bosonic 
and fermionic twist operators and calculates their correlations
functions.
The twist operators are located at the boundary of the world sheet.
Thus they are on the real axis when the world sheet is on the upper
half plane.
The bosonic twist operator $\sigma_\nu$ inserted at the origin 
changes the boundary conditions of the open string as shown in the
Figure 1.
The insertion of 
$\sigma_\nu$ changes the boundary conditions of the string to that of
a string joining the D0-brane to the D4-brane. 
The anti-twist operator $\sigma_{-\nu}$
changes the boundary conditions to that of a string joining the
D4-brane to the D0-brane.
From the mode expansions in \eq{bosemode1} and \eq{bosemode2} 
the bosonic twist operators $\sigma_\nu$ and $\sigma_{-\nu}$ have the
following OPE with the world sheet bosons
\FIGURE
{
\epsfig{file=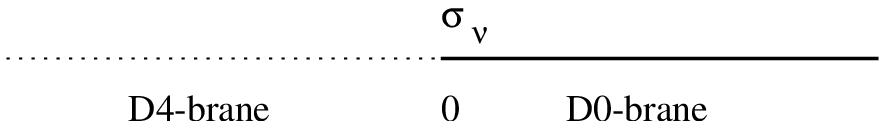}
\caption{Insertion of $\sigma_\nu$ at the origin changes the boundary
condition to that of a string joining the D0-brane and the D4-brane.}
}
\bea
\label{ope}
\del X^+(z) \sigma_\nu(w) = \frac{1}{(z-w)^{(1-\nu) } } 
\tau_\nu (w) & \; &
\del X^-(z) \sigma_\nu(w) = \frac{1}{(z-w)^\nu} \tau^\prime_\nu (w) \\
\nonumber
\del X^+(z) \sigma_{-\nu}(w) = 
\frac{1}{(z-w)^{\nu} } \tau_{-\nu} (w) & \; &
\del X^-(z) \sigma_{-\nu}(w) = \frac{1}{(z-w)^{(1-\nu)} }
\tau^\prime_{-\nu} (w) \\
\nonumber
\bar{\del} X^+(\bar{z}) \sigma_\nu(w) = 
-\frac{1}{(\bar{z}-w)^{1-\nu}} \tau_\nu (w) & \; &
\bar{\del} X^-(\bar{z}) \sigma_\nu(w) = 
-\frac{1}{(\bar{z}-w)^\nu} \tau^\prime_\nu (w) \\
\nonumber
\bar{\del} X^+(\bar{z}) \sigma_{-\nu}(w) = 
-\frac{1}{(\bar{z}-w)^{\nu} } \tau_{-\nu} (w) & \; &
\bar{\del} X^-(\bar{z}) \sigma_{-\nu}(w) = 
-\frac{1}{(\bar{z}-w)^{(1-\nu)}} 
\tau^\prime_{-\nu} (w) \\
\nonumber
\eea
$w$ is a point on the real axis. $\tau$'s are 
the excited twist operators.
We are interested in obtaining the following four point function of
the twist fields
\bea
Z_1(z_1, z_2, z_3, z_4) &=& 
\langle \sigma_{-\nu} (z_1) \sigma_\nu (z_2) \sigma_{-\nu}(z_3)
\sigma_\nu (z_4) \rangle \\ \nonumber
Z_2(z_1, z_2, z_3, z_4) &=& 
\langle \sigma_{1-\nu} (z_1) \sigma_{-(1-\nu} (z_2) 
\sigma_\nu(z_3)
\sigma_{-\nu} (z_4) \rangle
\eea
where $z_1, z_2, z_3, z_4$ are four points on the real axis. 
We will demonstrate the calculation for $Z_1$, the calculation of
$Z_2$ follows along similar lines.
We follow
the method in \cite{dfms} and as developed for open strings in
\cite{gns}. Consider the  auxiliary correlators
\be
g(z,w)=\frac{ \langle -\frac{1}{2} \del X^+(z) \del X^- (w) 
 \sigma_{-\nu} (z_1) \sigma_\nu (z_2) \sigma_{-\nu}(z_3)
\sigma_\nu (z_4) \rangle }
{\langle \sigma_{-\nu} (z_1) \sigma_\nu (z_2) \sigma_{-\nu}(z_3)
\sigma_\nu (z_4) \rangle}
\ee
and
\be
h(\bar{z},w)=\frac{ \langle -\frac{1}{2} \bar{\del} X^+(\bar{z}) 
\del X^- (w) 
 \sigma_{-\nu} (z_1) \sigma_\nu (z_2) \sigma_{-\nu(z_3)}
\sigma_\nu (z_4) \rangle }
{\langle \sigma_{-\nu} (z_1) \sigma_\nu (z_2) \sigma_{-\nu(z_3)}
\sigma_\nu (z_4) \rangle}
\ee
Define
\be
\omega_\nu(z) =
\frac{1}{[(z-z_1)(z-z_3)]^\nu}\frac{1}{[(z-z_2)(z-z_4)]^{1-\nu}}
\ee
Now $g(z, w)$ is given by
\bea
g(z, w) &=& \omega_\nu(z) \omega_{1-\nu} (w)  \left(
\nu\frac{(z-z_1)(z-z_3)(w-z_2)(w-z_4)}{(z-w)^2} \right.\\ \nonumber
&+& \left. (1-\nu) \frac{(z-z_2)(z-z_4)(w-z_1)(w-z_3)}{(z-w)^2} +
A(z_1,z_2,z_3, z_4) \right)
\eea
The form for $g(z,w)$ given  above has the required singularity
structure so that the \eq{ope} is obeyed.  When $z\rightarrow w$, then 
\be
g(z,w) = \frac{1}{(z-w)^2}
\ee
$h(\bar{z}, w)$ is given by
\bea
h(\bar{z}, w) &=& -\omega_\nu(\bar{z}) \omega_{1-\nu}(w) \left(
\nu\frac{(\bar{z}-z_1)(\bar{z}-z_3)(w-z_2)(w-z_4)}{(z-w)^2} \right. \\
\nonumber
&+&  \left.
(1-\nu) \frac{(\bar{z}-z_2)(\bar{z}-z_4)(w-z_1)(w-z_3)}{(z-w)^2} +
A(z_1,z_2,z_3, z_4) \right)
\eea
This form for $h(\bar{z}, w)$ also has the singularity structure so that
the \eq{ope} is obeyed. It  satisfies the condition
\be
h(\bar{z}, w) = - g(\bar{z}, w)
\ee
The origin of this condition can be seen from the mode expansions
\eq{bosemode1} and \eq{bosemode2}. The anti-holomorphic components 
$\bar{\partial} X^{\pm}(z)$ can be obtained from the holomorphic
components by replacing $z$ by $\bar{z}$ along with a negative sign.
We now determine $A$ from the monodromy conditions.
We have the monodromy condition \footnote{We are neglecting instanton
sectors as the D4-brane is extended in the spatial directions.}
\be
\label{mono}
\int_C g(z, w) dz + \int_C h(\bar{z}, w)d\bar{z} = 0
\ee
Where the contour $C$ is shown in Figure 2.
This monodromy condition arises because integration along the contour
$C$ gives the change in $X^+$ for a string  ending on the D0-brane at
the two ends. There are two equivalent nontrivial contours $C$ and
$C'$. 
\FIGURE{
\epsfig{file=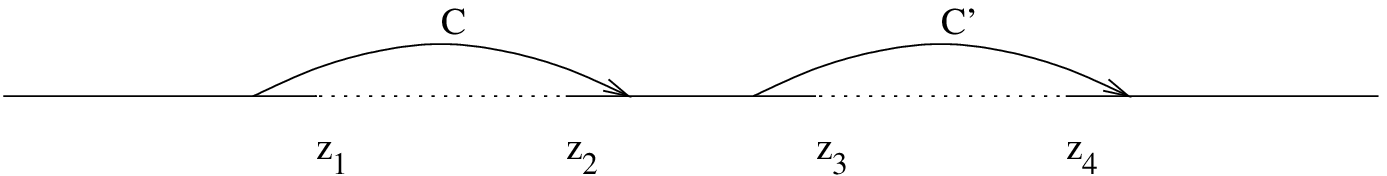}
\caption{Contours $C$ and $C'$ are equivalent nontrivial monodromy
contours}
}
To solve for $A$ first 
divide \eq{mono} by $\omega_{1-\nu}(w)$ and take $w\rightarrow\infty$. 
We then use the $SL(2,R)$ invariance to choose $z_1=0, z_2 = x,
z_3 =1$ and $z_4 =\infty$, where $0< x<1$. Then $A$ is given by
\be
A= -z_4 x(1-x) \frac{d}{dx}\ln F(\nu, 1-\nu , 1 ; x)
\ee
Here $z_4$ stands for $\infty$. $F(\nu, 1-\nu; x)$ is the
hypergeometric function whose integral representation is given by
\be
\label{hyper}
F(\nu, 1-\nu,1; x) = \frac{1}{\pi} \sin(\pi\nu) \int_0^1 dy
\frac{1}{ y^\nu (1-y)^{1-\nu} (1-xy)^\nu }
\ee

Now we can calculate the 
correlation function $Z_1(z_1, z_2, z_3, z_4)$. Consider 
the following correlation function
\bea
\frac{\langle 
T(z) \sigma_{-\nu }(z_1) \sigma_\nu (z_2) \sigma_{-\nu}(z_3) 
\sigma_\nu (z_4) \rangle} 
{ \langle \sigma_{-\nu }(z_1) \sigma_\nu (z_2) \sigma_{-\nu}(z_3) 
\sigma_\nu (z_4) \rangle}  &=& \lim_{w \rightarrow z} \left[ g(z, w)
- \frac{1}{(z-w)^2} \right] \\ \nonumber
&=& \frac{1}{2}\nu (1-\nu) \left( \frac{1}{(z-z_1)^2} +
\frac{1}{(z-z_3)^2}\right.  \\ \nonumber
&-& \left. \frac{1}{(z-z_3)^2} - \frac{1}{(z-z_4)^2}
\right) \\ \nonumber
&+&  \frac{A}{(z-z_1)(z-z_2)(z-z_3)(z-z_4)}
\eea
where $T$ is the stress energy tensor.
From the operator product
\be
T(z)\sigma_\nu(z_2) = \frac{1}{2}\frac{\nu(1-\nu) T(z_2)}{(z-z_2)^2} +
\frac{\partial_{z_2} \sigma_\nu (z_2)}{z-z_2 }
\ee
we can obtain the following differential equation for $Z_1(z_1, z_2,
z_3, z_4)$
\bea
\del_{z_2}\ln Z_1(z_1, z_2, z_3, z_4) 
&=& -\nu(1-\nu)\left( \frac{1}{(z_2 -z_1)} + \frac{1}{(z_2
-z_3)} -\frac{1}{(z_3-z_4)} \right)  \\ \nonumber
&+& \frac{A}{(z_2-z_1)(z_2-z_3)(z_2-z_4)}
\eea
Substituting the values of $z_1, z_2, z_3, z_4$ in the
above equation we obtain
\be
\partial_x Z_1(x) 
= -\nu(1-\nu) \left( \frac{1}{x} -\frac{1}{1-x}
\right) - \frac{d}{dx}\ln F(\nu, 1-\nu, 1; x)
\ee
This equation can easily be integrated to obtain
\be
Z_1(x)= \frac{1}{ [x(1-x)]^{\nu(1-\nu)} F(\nu, 1-\nu, 1, x) }
\ee
We now use the $SL(2,R)$ invariance to obtain $Z_1(z_1, z_2, z_3, z_4)$.
We get
\be
Z_1(z_1, z_2, z_3, z_4) = 
z_{21}^{-2h} z_{31}^{2h}
z_{41}^{-2h}z_{32}^{-2h} 
z_{42}^{2h} z_{43}^{-2h} 
\frac{1}{F(\nu, 1-\nu, 1; x)}
\ee
where $h = \nu (1-\nu)/2$, $z_{ij} = z_i-z_j$ and 
\be
x= \frac{z_{21}z_{43} }{ z_{31}z_{42} }
\ee
To construct the fermionic twist operators we first bosonize the
fermions by
\be
\psi^+(w) = e^{iH(w)} \;\;\;\; \psi^-(w) = e^{-iH(w)}
\ee
where $H$ is a free boson.
The fermionic  twist operator are is given by $e^{i\nu H}$ and the
anti-twist operator is given by $e^{-i\nu H}$.
The correlation function for the fermionic twists are easy to evaluate
and they are given by
\be
\langle e^{-i\nu H(z_1)}e^{i\nu H(z_2)}e^{-i\nu H(z_3)}e^{-i\nu H(z_4)}
\rangle 
= 
z_{21}^{-\nu^2} z_{31}^{\nu^2}
z_{41}^{-\nu^2}z_{32}^{-\nu^2} z_{42}^{\nu^2}
z_{43}^{-\nu^2}
\ee
Putting the bosonic and the fermionic twists together one obtains
\be
\label{cortwist}
 \langle \Sigma_{-\nu}(z_1)\Sigma_{\nu}(z_2)
\Sigma_{-\nu}(z_3)\Sigma_{\nu}(z_4) \rangle 
= z_{21}^{-\nu} z_{31}^{\nu}
z_{41}^{-\nu}z_{32}^{-\nu} z_{42}^{\nu}
z_{43}^{-\nu} \frac{1}{F(\nu, 1-\nu, 1, x)}
\ee
Using similar methods we obtain the correlation function $Z_2(z_1,z_2,
z_3, z_4)$. It is given by
\bea
Z_2(z_1, z_2, z_3, z_4) = 
z_{21}^{-2h} z_{31}^{-2h}
z_{41}^{2h}z_{32}^{2h} 
z_{42}^{-2h} z_{43}^{-2h} 
\frac{1}{F(\nu, 1-\nu, 1; y)}
\eea
Here $y = x/(x-1)$. Using $Z_2$ we can find the following correlation
functions
\bea
\label{4pt2}
\langle
\Sigma_{1-\nu}(z_1)\Sigma_{-(1-\nu)}(z_2)
\Sigma_{\nu}(z_3)\Sigma_{-\nu}(z_4)
\rangle &=& \frac{1}{z_{21}^{1-\nu}
z_{43}^\nu} \frac{1}{F(\nu, 1-\nu, 1;y)}
\\ \nonumber
\langle
\Sigma_{-\nu}(z_1)\Sigma_{\nu}(z_2)
\Sigma_{-(1-\nu)}(z_3)\Sigma_{1-\nu}(z_4)
\rangle &=& \frac{1}{z_{21}^{\nu}
z_{43}^{1-\nu} } \frac{1}{F(\nu, 1-\nu, 1;y)} \nonumber
\eea
For completeness we write down the two point function of the twist
operators. This is fixed by conformal invariance.
\be
\label{2pt}
\langle \Sigma_\nu (z_1) \Sigma_{-\nu} (z_2) \rangle = 
\frac{1}{(z_2 -z_1)^\nu}
\ee
We have chosen a normalization for the two point function which
differs from the conventional one to make sure all coefficients are
real in the tachyon potential. The four point function of the twist
operators in \eq{cortwist} and \eq{4pt2}
are consistent with this normalization of the two point
function \footnote{If we had worked with the conventional
normalization we can ensure that the coefficients of the tachyon
potential are real by a redefinition of the tachyon field.}.

\section{Details on the calculation of the tachyon potential}

The vertex operators $\hat{T}$ and $\hat{E}$ are all primary,
therefore they transform under a conformal transformation $f$ as
\be
f\circ {\cal O}(0) = (f^{\prime}(0))^h {\cal O}(0)
\ee
Here ${\cal O}$ is a primary operator of dimension $h$.

\subsection{The quadratic term}

We now focus on the evaluation of the quadratic term in the string
field theory action given by
\be
S^{(2)} = \frac{1}{4g^2} \langle\langle (\hat{Q}_B\hat{\Phi})
(\hat{\eta}_0 \hat{\Phi}) \rangle\rangle
\ee
Substituting $\hat{\Phi}= \hat{T} + \hat{E}$, the terms which
contribute in $S^{(2)}$ are
\be
S^{2} = \frac{1}{4g^2}
\left( 
\langle\langle (\hat{Q}_B\hat{T})
(\hat{\eta}_0\hat{T}) \rangle\rangle 
+ \langle\langle (\hat{Q}_B\hat{E})
(\hat{\eta}_0\hat{E}) \rangle\rangle  \right)
\ee
The cross terms do not contribute due to twist conservation.
$\hat{T}$ is an operator of dimension $-(1-(\nu +\nu'))/2$ 
and $\hat{E}$ is an operator of dimention $(1-(\nu + \nu'))/2$.
We  calculate the correlations functions on the upper half plane.
Using \eq{2pt} and evaluating the 
traces over the Chan-Paton factors and grouping the terms we obtain
\bea
S^{(2)} &=& \frac{1}{g^2} \left(
+\frac{1}{2}(1- (\nu + \nu'))
z_{21}^{1-(\nu +\nu')} (g^{(2) \prime}_1 (0)
g^{(2)\prime}_2 (0)) ^{-(1-(\nu + \nu')/2}t_-t_+ \right.
\\ \nonumber
&-&\left. \frac{1}{2}(1- (\nu + \nu'))
z_{21}^{-1+(\nu +\nu')} (g^{(2)\prime}_1 g^{(2)\prime}_2) 
^{(1-(\nu + \nu')/2}e_-e_+
\right)
\eea
where $z_1 = g^{(2)}_1 (0)$ and $z_2 = g^{(2)}_2 (0)$. The $g$'s are
defined in \cite{bsz}. We write them down here for completeness
\bea
g_1^{(2)} = \tan(-\frac{\pi}{4}) \;&\;&\; g_1^{(2)\prime} = 2
\\ \nonumber
g_2^{(2)} = \tan(\frac{\pi}{4}) \;&\;&\; g_1^{(2)\prime} = 2
\eea
Substituting the values of each of the terms in
$S^{(2)}$ we obtain
\be
S^{(2)}= \frac{1}{g^2}\left( \frac{1}{2}(1-(\nu+\nu'))t_-t_+
-\frac{1}{2}(1-(\nu+\nu'))e_-e_+ \right)
\ee
\subsection{The cubic term}

The cubic term in the superstring field theory action 
is given by
\be
S^{(3)} = \frac{1}{12 g^2} \langle\langle (\hat{Q}_B
\hat{\Phi})(\hat{\Phi} (\hat{\eta}_0 \hat{\Phi} ) - (\hat{\eta}_0
\hat{\Phi}\hat{\Phi)} \rangle\rangle
\ee
Substituting $\hat{\Phi} = \hat{T} + \hat{E}$ in $S^{(3)}$ we see that
it vanishes. This 
can be easily seen by taking the traces of the external 
Chan-Paton factors.

\subsection{The quartic term}

The quartic term in the string field theory action is given by
\be
S^{(4)} = \frac{1}{48g^2} \langle\langle (
\hat{Q}_B\hat{\Phi}) (
\hat{\Phi}^2(\hat{\eta}_0\hat{\Phi}) - 2
\hat{\Phi}(\hat{\eta}_0\hat{\Phi}) + (\hat{\eta}_0\hat{\Phi})
\hat{\Phi}^2 ) \rangle\rangle
\ee
From the fact that there should be only three $c$ ghosts for the
correlation functions to contribute we can see that
after the  substitution 
$\hat{\Phi} = \hat{T} + \hat{E}$
there is a factorization of the action given by
\be
S^{(4)} = {\cal S}\times{\cal T} 
\ee
where
\bea
{\cal S} &=& \frac{1}{48g^2}\langle\langle 
(Q_B\otimes\sigma_3 O\otimes\sigma_1) \left(
( O\otimes\sigma_1)^2(\eta_0\otimes\sigma_3 O\otimes\sigma_1)
\right. \\ \nonumber
&-& 2\left. O\otimes\sigma_1 (\eta_0 \otimes\sigma_3O\otimes\sigma_1)
+ (\eta_{0}\otimes\sigma_3 O\otimes\sigma_1) (O\otimes\sigma_1)^2
\right)\rangle\rangle   \\ \nonumber
{\cal T} &=& \langle\langle PPPP\rangle\rangle
\eea
Here $O = \xi c e^{-\phi}$ and
\bea
P &=& 
t_+ \Sigma_\nu \Sigma_{\nu '}
 \otimes \sigma_+
 + t_- \Sigma_{-\nu} \Sigma_{-\nu'} 
 \otimes \sigma_-  \\ \nonumber
&+& e_+ \Sigma_{1-\nu} \Sigma_{1-\nu '}
 \otimes \sigma_+
 + e_- \Sigma_{-(1-\nu)} \Sigma_{-(1-\nu')} 
 \otimes \sigma_- 
\eea
The value of ${\cal S}$ has been calculated in \cite{berk2, bsz}.
${\cal S} = -1/(2g^2)$. 
We have to use the correlation functions found 
in Appendix A to evaluate ${\cal T}$.
As an example we write down the coefficient of $(t_-t_+)^2$
\bea
\label{coef}
 2 
\left( g_1^{(4)\prime}(0) 
g_2^{(4)\prime}(0) 
g_3^{(4)\prime}(0) 
g_3^{(4)\prime}(0) 
\right)^{(\nu+\nu')/2} \\ \nonumber
\langle 
\Sigma_{-\nu}(z_1)
\Sigma_{\nu}(z_2)
\Sigma_{-\nu}(z_3)
\Sigma_{\nu}(z_4) \rangle
\langle \Sigma_{-\nu'}(z_1)
\Sigma_{\nu'}(z_2)
\Sigma_{-\nu'}(z_3)
\Sigma_{\nu'}(z_4) \rangle
\eea
where
$ z_1 = g_1^{(4)}(0)
z_2 = g_2^{(4)}(0)
z_3 = g_3^{(4)}(0)
z_4 = g_4^{(4)}(0)$. 
and the $g$'s are defined in \cite{bsz}
They are given by
\bea
g_1^{(4)}(z) &=& -4 + 6z -9z^2 + \cdots \\ \nonumber
g_2^{(4)}(z) &=& -1 + \frac{3}{4}z -\frac{3}{16}z^2 + \cdots \\
\nonumber
g_3^{(4)} (z) &=& 0 + \frac{2}{3}z +\frac{1}{9}z^2 + \cdots \\
\nonumber
g_4^{(4)}(z) &=& 2 + 3z + 3z^2 + \cdots \\
\eea
Substituting these values 
in \eq{coef} and using the correlation
function in \eq{cortwist} we find that the coefficient is that given 
in \eq{tpot}. Note that the value of argument $x$ of the
hypergeometric function is $1/2$ and $y=-1$.

\end{document}